\documentclass{article}
\usepackage{amsmath,amsthm}
\usepackage{amssymb,latexsym}
\usepackage[mathscr]{eucal}
\usepackage{setspace}
\usepackage{graphics}
\usepackage{array}

\newcommand{\be}{\begin{equation}}
\newcommand{\ee}{\end{equation}}

\newcommand{\pr}{\prime}
\newcommand{\al}{\alpha}

\newcommand{\rp}{\right)}
\newcommand{\lp}{\left(}
\newcommand{\rb}{\right]}
\newcommand{\lb}{\left[}

\usepackage{graphicx}
\graphicspath{{converted_graphics/}}
\begin{document}

\title{\bf Compressible Turbulence:~Multi-fractal Scaling in the Transition to the Dissipative Regime}         
\author{B.K. Shivamoggi\footnote{\large {Permanent Address: University of Central Florida, Orlando, FL 32816-1364}} \\
Los Alamos National Laboratory \\
Los Alamos, NM 87545}        
\date{}
\maketitle

\large{\bf Abstract}

Multi-fractal scaling in the transition to the dissipative regime for fully-developed compressible turbulence is considered. The multi-fractal power law scaling behavior breaks down for very small length scales thanks to viscous effects. However, the effect of compressibility is found to extend the single-scaling multi-fractal regime further into the dissipative range. In the ultimate compressibility limit, thanks to the shock waves which are the appropriate dissipative structures, the single-scaling regime is found to extend indeed all the way into the full viscous regime. This result appears to be consistent with the physical fact that vortices stretch stronger in a compressible fluid hence postponing viscous intervention. The consequent generation of enhanced velocity gradients in a compressible fluid appears to provide an underlying physical basis for the previous results indicating that fully-developed compressible turbulence is effectively more dissipative than its incompressible counterpart.

\pagebreak

\section{Introduction}       

Compressibility effects on fully-developed turbulence (FDT) are of importance in modern technological flow processes like makes of supersonic projectiles, hypersonic re-entry vehicles and high-speed internal flows in gas-turbine engines. Astrophysical processes like star formation in self-gravitating dense interstellar gas clouds via Jeans' instability (Chandrasekhar \cite{chan} and Shu \cite{shu}) are other cases in point. So, considerable effort has been directed on this problem to date (Moiseev et al. \cite{moi}, Passot and Pouquet \cite{pas1}, \cite{pas2}, Porter et al. \cite{por}, Erlebacher et al. \cite{erl}, Shivamoggi \cite{shi13}-\cite{shi5} and Kritsuk et al. \cite{kri17}-\cite{kri18}).

The multi-fractal model was used by Shivamoggi~\cite{shi1}-\cite{shi3}, \cite{shi4}, \cite{shi5} to describe the numerically observed spatial intermittency in fully-developed compressible turbulence (Passot and Pouquet \cite{pas1}, \cite{pas2} and Lee et al. \cite{lee}). A multi-fractal object exhibits a global scaling structure that is described by a continuous spectrum of scaling exponents $\al$ and in a certain range $I\equiv [\al_{min}, \al_{max}]$. Each $\al \in I$ has the support set $S(\al)\subset \mathbb{R}^3$ of fractal dimension $f(\al)$ such that the velocity increment over a small distance $\ell$ has the scaling behavior (Frisch and Parisi~\cite{fripar}),
\be\tag{1}
\delta v (\ell) \sim \ell^{\al},\; \ell  \text{small}.
\ee
The sets $S(\al)$ are nested so that $S(\al^\pr) \subset S (\al)$ for $\al^\pr < \al$. The fractal dimension $f(\al)$~is obtained via a Legendre transformation of the scaling exponent $\xi_p$ of the $p th$-order structure function, 
\be\tag{2}
S_p (\ell) \equiv~ <\mid\delta v (\ell) \mid ^p >~ = \int\limits_{\al_{min}}^{\al_{max}}d \mu(\al) \ell^{\al p + 3 - f(\al)}\sim \ell^{\xi_p}
\ee
according to 
\be\tag{3}
\xi_p = \inf_\al [\al p + 3 - f(\al)],~\text{say for}~ \al = \al_*.
\ee
where,
\be\tag{4}
\frac{df (\al^*)}{d \al} = p.
\ee
The factor $(3-f(\al))$ is the fraction of the coarse-graining measure of the volume of the set $S(\alpha)$ over the total volume.

When viscous dissipative effects arise, the multi-fractal power law described by equation (2) breaks down for very small $\ell$ and the cut-offs are determined by external parameters like the Reynolds number. In such cases, as in incompressible turbulence (Wu et al. \cite{wu}, Frisch et al. \cite{fri}), the multi-fractal in question turns out to exhibit a pseudo-algebraic behavior with certain universal features, albeit on using a suitable rescaling via a multi-scaling method (Wu et al. \cite{wu}). Similar situation arises also for a passive scalar diffusing in a random velocity field (Shivamoggi \cite{shi12}).

\section{Multi-fractal Scaling in the Inertial Range: Behavior in the Ultimate Compressibility Limit}

In order to relate the fractal dimension $f (\al)$ to the generalized fractal dimension $D_q$ of the kinetic energy dissipation field $\hat{\varepsilon}$, one considers a coarse-grained probability measure given by the total kinetic energy dissipation ocurring in a box size $\ell$ -
\be\tag{5}
E(\ell) \sim \hat{\varepsilon} (\ell) \ell^3
\ee
and then covers the support of the measure $\hat{\varepsilon}$ with boxes of size $\ell$ and sums the moments $[E (\ell)]^q$ over all boxes. Noting the asymptotic scaling behavior of these moments (Halsey et al. \cite{hal})
\be\tag{6a}
\sum_{i = 1}^{N (\ell)} [E (\ell)]^q \sim \ell^{(q - 1) D_q}
\ee
or
\be\tag{6b}
\int d \mu (\al) \ell^{\lb \lp \frac{3 \gamma - 1}{\gamma - 1} \rp \al + 2 \rb q - f (\al)} \sim \ell^{(q - 1) D_q}.
\ee
$\gamma$ is the ratio of specific heats of the fluid and may be used as a compressibility parameter $(1 < \gamma < \infty)$ - the incompressible fluid corresponds to the limit $\gamma \Rightarrow \infty$ while the ultimate compressibility case given by the limit $\gamma \Rightarrow 1$ corresponds to Burgers turbulence.

The dominant terms in the integral in (6)b, in the limit of small $\ell$, may again be extracted using the method of steepest descent:
\be\tag{7}
\lb \lp \frac{3 \gamma - 1}{\gamma - 1} \rp \al^* + 2 \rb q - f (\al^*) = (q - 1) D_q
\ee
with,
\be\tag{8}
\frac{d f (\al^*)}{d \al} = \lp \frac{3 \gamma - 1}{\gamma - 1} \rp q.
\ee

Eliminating $f (\al)$ and putting $q = \lp \frac{\gamma - 1}{3 \gamma - 1} \rp p$, we obtain from (3) and (7) (Shivamoggi \cite{shi13}, \cite{shi1}),
\be\tag{9}
\xi_p = \lp \frac{\gamma - 1}{3 \gamma - 1} \rp p - \frac{1}{3} \lb \lp \frac{3 \gamma - 3}{3 \gamma - 1} \rp p - 3 \rb \lb 3 - D_{\lp \frac{\gamma - 1}{3 \gamma - 1} \rp p} \rb.
\ee

On noting the scaling behavior the density $\rho$ (Shivamoggi \cite{shi13}) (which follows on applying scale-invariance arguments directly to the Navier-Stokes equations for a compressible fluid in conjunction with the scale-invariance condition on the kinetic energy dissipation field),
\be\tag{10}
\rho \sim (\delta v)^{\frac{2}{\gamma - 1}}
\ee
the scaling behavior of the kinetic energy is given by
\be\tag{11}
<\rho (\delta v)^2> ~\sim ~<\delta v>^{\frac{2 \gamma}{\gamma - 1}}.
\ee
(11), in conjunction with (2), leads to
\be\tag{12}
<\rho (\delta v)^2> ~\sim ~\ell^{\xi_{\lp \frac{2 \gamma}{\gamma - 1} \rp}}
\ee
where, we have, from (9),
\be\tag{13}
\xi_{\lp \frac{2 \gamma}{\gamma - 1} \rp} = \frac{2 \gamma}{3 \gamma - 1} + \lp \frac{\gamma - 1}{3 \gamma - 1} \rp \lb 3 - D_{\lp \frac{2 \gamma}{3 \gamma - 1} \rp} \rb.
\ee

In the ultimate compressible limit $\gamma \Rightarrow 1$, (12) yields
\be\tag{14}
<\rho (\delta v)^2> ~\sim ~\ell
\ee
which leads to the Kadomtsev-Petviashvili \cite{kad} spectral law for compressible turbulence -
\be\tag{15}
E(k) \sim k^{-2}.
\ee

On the other hand, for a fractally homogeneous turbulence, we have
\be\tag{16}
D_q = D_0, \forall q
\ee
so (9) becomes
\be\tag{17}
\xi_p = \lp \frac{\gamma - 1}{3 \gamma - 1} \rp p - \frac{1}{3} \lb \lp \frac{3 \gamma - 3}{3 \gamma - 1} \rp p - 3 \rb (3 - D_0).
\ee

If we identify the dissipative structures in compressible turbulence with shock waves, we have $D_0 = 2$, and (17) yields
\be\tag{18}
\xi_p = 1, \forall p.
\ee
(12), in conjunction with (18), leads to
\be\tag{19}
<\rho (\delta v)^2> ~\sim ~\ell
\ee
in agreement with (14).

\section{Multi-fractal Scaling in the Transition to the Dissipative Regime}

The dissipative effects materialize when typically the eddy turn-over time exceeds the viscous diffusion time, i.e., 
\be\tag{20}
\frac{\ell}{\delta v} > \frac{\ell^2}{\mu}
\ee
or
\be\tag{21}
\delta v < \frac{\mu}{\ell} \equiv u \sim \ell^{\overline\al}.
\ee
(21) implies the presence of a viscous cut-off $u$ so that the boxes with measures $\delta v < u ~$or indices $\al > \overline\al$, where
\be\tag{22}
\overline\al = \frac{\ell n u}{\ell n \ell} = -1 -\frac{\ell n \mu}{\ell n(\frac{1}{\ell})}
\ee
would be empty and not contribute therefore to the integral for $S_p$ in (2). Note that different length scales (below the Kolmogorov microscale $\eta$) have different viscous cut-offs. Thus, 
\be\tag{23}
S_p (\ell, u) = \int\limits_{\al_{min}}^{\overline\al} d \mu (\al)\ell^{\al p + 3 - f(\al)}
\ee
which signifies the existence of a continuous range of viscous cut-offs in the multi-fractal model for the inertial range. As these cut-offs vary, a cross-over in the scaling of the structure function appears for $\ell\sim\eta$ (Wu et al. \cite{wu}).

For compressible turbulence, noting that
\be\tag{24}
\hat{\varepsilon}\sim \frac{\rho v^3}{\eta} \sim \mu \frac{v^2}{\eta^2}
\ee
we have,
\be\tag{25}
\eta \sim \frac{\mu}{\rho v}~ \text{and}~ v \sim \lp \frac{\hat{\varepsilon}\eta}{\rho} \rp ^{\frac{1}{3}}
\ee
from which, 
\be\tag{26}
\eta \sim \lp \frac{\mu^3}{\rho^2 \hat{\varepsilon}}\rp^{\frac{1}{4}}.
\ee

On the other hand, on noting the following scaling behavior (Shivamoggi \cite{shi13}),
\be\tag{27}
\rho (\ell) \sim \ell^{\frac{2}{3\gamma-1}}
\ee
and using (26), we have
\be\tag{28}
\ell > \eta : ~\ell^{\frac{3\gamma}{3\gamma-1}} > \mu^{\frac{3}{4}}~  \text{  or} ~ -\frac{\ell n \mu}{\ell n {\lp\frac{1}{\ell}\rp }}> \frac{4\gamma}{3\gamma-1}.
\ee

Using (28), we have from (22), 
\be\tag{29}
\ell > \eta: ~\overline\al > \frac{\gamma +1}{3\gamma-1} > \al_*
\ee
where $\al_*$ is defined in equation (3) and gives the leading order contribution to the scaling exponent of the structure function. So, in this interval, we have the usual single-scaling multi-fractal regime in the inertial range. On noting that
\be\tag{30}
\frac{\gamma+1}{3\gamma-1} = \frac{1}{3} + \frac{\frac{4}{3}}{3\gamma-1}
\ee
we observe, from (29), that the effect of compressibility is to extend the single-scaling multi-fractal regime further into the dissipative range, provided of course $\al_*<\overline\al$.

Further, in the ultimate compressibility limit $\gamma \Rightarrow 1$,
\be\tag{31}
\frac{\gamma + 1}{3 \gamma - 1} \Rightarrow 1
\ee
so, according to (29), at the onset of viscous effects, we have
\be\tag{32}
\delta v \sim \ell
\ee
which is, however, the scaling signifying a smooth velocity field appropriate for the full viscous regime. So, in the ultimate compressibility limit, thanks to the shock waves which are the appropriate dissipative structures, the single-scaling regime extends indeed all the way into the full viscous regime.

On the other hand, when $\ell$ falls well below $\eta$ so that $\overline\al < \al_* < (\frac{\gamma+1}{3\gamma-1})$, the minimum of $\xi_p$ is attained for $\al = \overline\al$. We now have a pseudo-algebraic behavior (as in the incompressible case  (Jensen et al. \cite{jen})) induced by dissipative effects - the power law in question has a scale-dependent exponent, albeit slowly varying via proportionality to $\ell n \ell$.

Thus, for small $\ell$, we have
\be\tag{33}
S_p(\ell, u)\sim
\begin{cases}
\ell^{\xi_p} & ,\text{if} ~\overline\al > \al_*\\
\\
\ell^{\overline\al p+3-f(\overline\al)}&, \text{if}~ \overline\al\leq\al_*.
\end{cases}
\ee
It may be noted that the saddle-point evaluation method used in deriving (33) becomes exact in the limit $\overline\al \Rightarrow \al_{min}$ when one obtains the result corresponding to a homogeneous fractal. 

It is possible to bring out universal features even in a non-power-law situation, albeit on using a suitable rescaling associated with multi-scaling reflecting variations in the cut-off parameter $u$ by introducing
\be\tag{34}
F_p(\theta) \equiv \frac{\ell n S_p}{\ell n u},~ \theta \equiv \frac{\ell n \ell}{\ell n u} = \frac{1}{\overline\al}.
\ee
(33) then implies
\be\tag{35}
F_p (\theta) =
\begin{cases}
\theta\xi_p,&~\text{if}~\theta < \frac{1}{\al_*}\\
\\
p + \theta [3- f(\frac{1}{\theta})],&~ \text{if}~\theta\geq\frac{1}{\al_*}.
\end{cases}
\ee
The first regime in (35) corresponds to the single-scaling regime where the graph of $F_p$ vs. $\theta$ is a straight line with slope $\xi_p$. The second regime in (35) has the structure function data for different Reynolds numbers collapsing onto a single curve in the neighborhood of the viscous regime - multi-fractal universality for compressible turbulence.

\section{Discussion}

In fully-developed compressible turbulence, when the dissipative effects arise, the multi-fractal power-law scaling behavior breaks down for very small length scales and the cut-offs are determined by external parameters like the Reynolds number. However, if one uses a suitable rescaling, the multi-fractal in question is shown still to exhibit a multi-scaling behavior with certain universal features, as in the incompressible case, via the existence of a continuous range of viscous cut-offs. The effect of compressibility is found to extend the single-scaling multi-fractal regime further into the dissipative range - it is as though compressibility allows length scales to become smaller without forcing viscous dissipative effects to become operational. In the ultimate compressibility limit, thanks to the shock waves which are the appropriate dissipative structures, the single-scaling regime is found to extend indeed all the way into the full viscous regime. This result appears to be consistent with the fact that vortices stretch stronger in a compressible fluid hence postponing viscous intervention. The consequent generation of enhanced velocity gradients in a compressible fluid appears to provide an underlying physical basis for the results (via group-theoretical arguments applied to a Hopf-type functional equation formulation (Moiseev et al. \cite{moi}), scaling arguments applied directly to the compressible Navier-Stokes equations (Shivamoggi \cite{shi13}, \cite{shi1}), direct numerical simulations (Passot and Pouquet \cite{pas2}, Erlebacher et al. \cite{erl}, and Kritsuk et al. \cite{kri17}, \cite{kri18}) indicating that fully-developed compressible turbulence has steeper energy spectra and is effectively more dissipative than its incompressible counterpart. On the other hand, the enhanced vortex stretching is consistent with the results that - 
\begin{itemize}
  \item the velocity-space singularities are stronger in fully-developed compressible turbulence;
  \item the mean kinetic energy of a typical spectral mode in an equilibrium distribution is reduced by compressibility effects (Shivamoggi \cite{shi19}).
\end{itemize}

\end{document}